\documentstyle[aps,amssymb,twocolumn]{revtex}

\input{tcilatex}

\begin{document}
\title{Angular Dependence of the Nonlinear Transverse Magnetic Moment of YBa$_{2}$Cu%
$_{3}$O$_{6.95}$ in the Meissner State.}
\author{Anand Bhattacharya, Igor Zutic, O.T.Valls and A.M.Goldman}
\address{School of Physics and Astronomy, University of Minnesota, Minneapolis, MN\\
55455, USA}
\author{Ulrich Welp and Boyd Veal}
\address{Materials Science Division, Argonne National Laboratory, Argonne, Illinois
60439.}
\date{Sept. 25, 1998}
\maketitle
\title{}
\date{The Date }

\begin{abstract}
The angular dependence of the nonlinear transverse magnetic moment of
untwinned high-quality single crystals of YBa$_{2}$Cu$_{3}$O$_{6.95}$ has
been studied at a temperature of 2.5K using a low frequency AC technique.
The absence of any signature at angular period {\em 2}${\em \pi }${\em /4 }%
is analyzed in light of the numerical predictions of such a signal for a
pure {\em d}$_{{\em x}^{2}\text{-}{\em y}^{2}}$ order parameter with line
nodes. Implications of this null result for the existence of a non-zero gap
at all angles on the Fermi surface are discussed.
\end{abstract}

\begin{abstract}
\end{abstract}



Measurements of the low energy quasiparticle excitation spectrum of high
temperature superconductors (HTS) are crucial in elucidating the nature of
the pairing state and gap function in these materials. For a {\it d}$%
_{x^{2}-y^{2}}$ order parameter, the gap function varies as ${\it \Delta }$($%
{\it \varphi }$) ={\it \ 
\mbox{$\vert$}%
}${\it \Delta }_{{\it 0}}${\it cos(2}${\it \varphi }${\it ) 
\mbox{$\vert$}%
}, riding on a cylindrical Fermi surface in momentum space, with zeros or
line nodes at ${\it \varphi }${\it \ = (}${\it \pi }${\it /4 + n}${\it \pi }$%
{\it /2)}. Here,{\it \ }${\it \Delta }_{{\it 0}}$ is about 25meV \cite{photo}
for YBa$_{2}$Cu$_{3}$O$_{6.95}$ (YBCO), and ${\it \varphi }$ is the angle
measured from the crystallographic {\it a} or {\it b} directions. The
density of states for quasiparticle excitations near these nodes on the
Fermi surface increases linearly with energy. In a two fluid model, this
results in the superconducting condensate density {\it n}$_{s}${\it (T)}
decreasing linearly with increasing temperature from its zero temperature
value {\it n}$_{s}${\it (0)} \cite{annett}. In conventional {\em s-wave }%
superconductors, there is a non-zero gap everywhere on the Fermi surface for
quasiparticle excitations, and the depletion of the condensate is
exponentially small at the lowest temperatures

The temperature dependence of the in-plane London penetration depth \cite
{hardy93}{\it \ }${\it \lambda }${\it (T)} in these materials has been
widely cited as evidence in support of the presence of line nodes in the gap
function, though at the lowest temperatures there is some deviation from the
expected linear behavior, even in the best samples \cite{hardy98}.
Measurements of the magnetic field dependence of the low temperature
specific heat have also been interpreted as evidence for the presence of
nodes, but the analysis is subject to the limitations of a many-parameter fit%
\cite{spheat1}. Data from later experiments, taken over a wider range of
temperature and field do not yield the same conclusions upon analysis \cite
{spheat2}. Recent studies on the scaling of specific heat data has also been
cited as evidence in support of nodes \cite{scaling}, but the scaling works
poorly, and only over a limited temperature range. Amongst experiments that
provide angular information, angle resolved photemission experiments (ARPES)
can be interpreted as showing a minimum in the energy gap in the (110)
direction in YBCO and BSCCO \cite{photo,shen}, but have a resolution of only
a few meV (1meV = 11.62K) and thus cannot resolve excitations or an
underlying {\em s-wave} gap at the lowest energies. Inelastic neutron
scattering shows evidence in support of minima \cite{neutrons1}, but there
too, the resolution is 1meV at best. Thus, one can safely say that the
overall picture is quite ambiguous regarding the possibility of an
underlying gap below 1meV.

In this Letter, we describe an experimental technique for distinguishing
nodes from deep minima (quasinodes) in the order parameter, that also
provides information about the angular position of these nodes or
quasinodes, by probing the existence of low lying excitations in response to
an applied magnetic field in the Meissner regime. Here, the kinetic energy
of the superflow of the screening currents provides the energy for
quasiparticle excitations. Our null result for this probe rules out the
existence of nodes, and allowing for quasinodes, sets a lower bound on the
size of the underlying gap.

For a type II superconductor in a magnetic field in the Meissner regime,
screening supercurrents flow in a volume near the surface given
approximately by the penetration depth ${\em \lambda }$. For the condensate
participating in these currents, the quasiparticle excitation spectrum is
modified by a semiclassical `Doppler shift' to {\it E(k) = }$\sqrt{(\Delta
_{k}^{2}+\epsilon _{k}^{2})}${\it \ + }$\widehat{{\it v}}_{s}\cdot \widehat{%
{\it v}}_{F}$ \cite{bardeen}, where{\bf \ }$\widehat{{\it v}}_{s}$ is the
superfluid flow field and $\widehat{{\it v}}_{F}$ is the Fermi velocity. For
a gap function with nodes, this leads to quasiparticle excitations even at
zero temperature \cite{ys}. Due to the linear density of states near a line
node, the depletion of the condensate due to quasiparticles is proportional
to ($\widehat{{\it v}}_{s}\cdot \widehat{{\it v}}_{F}$)$^{2}$. These
quasiparticles create a `backflow' which is then responsible for a nonlinear
contribution to the magnetization that goes as {\it H}$^{2}{\it \lambda }%
^{2} $, where $\widehat{{\it H}}$ is the applied magnetic field, and {\it v}$%
_{s}${\it \ }$\sim $ {\it H}${\it \lambda }$. In the case of a {\it d}$%
_{x^{2}-y^{2}}$ order parameter, the gap function has a four-fold angular
symmetry, and this gives rise to an intrinsic nonlinear {\em transverse}
magnetization, superimposed on the nominal diamagnetic response\cite{zv1}.
Since this effect is felt only by a fraction of the condensate within a
thickness of ${\it \lambda }$ of the surface, the nonlinear magnetic moment
is proportional to the surface area of the sample, and not the full sample
volume. Also, since ${\it \lambda }$ 
\mbox{$>$}%
\mbox{$>$}%
${\it \xi }$, where ${\it \xi }$ is the in-plane coherence length, this is a
bulk effect.

We are interested in the nature of the gap function for supercurrents
flowing within the Cu-O planes of the superconductor, which consists of
layers of these planes. The YBCO crystals are flat with the {\it c}-axis
oriented perpendicular to the crystal plane which is the {\em a-b} plane.
The magnetic field is applied parallel to the crystal plane and the
transverse magnetic moment is measured in the crystal plane perpendicular to
the applied field. As the crystal is rotated in the applied magnetic field
about the {\it c}-axis, the screening currents flow in different directions
relative to the in-plane gap function, which is pinned to the crystal
lattice. This leads to an angular modulation of the nonlinear transverse
magnetic moment that provides {\it angle-resolved} information about the low
lying excitations in YBCO.

Numerical calculations of the nonlinear transverse magnetic moment{\it \ }(%
{\em m}$_{T}$) of a sample with a finite disk shaped geometry have been
carried out for pure {\it d-wave} and mixed order parameters. For pure {\em %
d-wave}, the calculation predicts the amplitude of the expected four-fold
modulation of {\em m}$_{T}$ with angular period ${\em \pi /2}$ \cite{zv1}, a
consequence of the symmetry of the gap and angular position of the nodes.
The presence of a small {\em s-wave} component as in {\it d+s} changes the
angular position of the line nodes, and introduces a component in the
angular modulation of {\em m}$_{T}$ with period ${\em \pi }$ \cite{zv2}.
However, the angular period ${\em \pi /2}$ component is not adversely
affected by small additions of {\it s-wave}. Anisotropy of the Fermi surface
and anisotropy in ${\em \lambda }_{a}$ and ${\em \lambda }_{b}$ have also
been considered, and have very similar consequences. For experimentally
determined values of such anisotropy\cite{abaniso}, there is a small
additional angular period ${\em \pi }$ component, but the period ${\em \pi /2%
}$ component is essentially unchanged.{\em \ }On the other hand, for a{\it \
d-wave} like order parameter, with varying levels of a non-zero quasinode,
as in {\it d}$_{x^{2}-y^{2}}$+ i{\it s} and {\it d}$_{x^{2}-y^{2}}$ + i{\it d%
}$_{xy}$ symmetries, the amplitude of the period ${\em \pi /2}$ component is
suppressed (see Fig.2 inset). In addition, to prevent the nonlinear Meissner
effect (NLME) from being thermally washed out, the temperature has to be
such that $\frac{T}{{\it \Delta }_{0}}$ 
\mbox{$<$}%
$\frac{H}{H_{o}}$, where {\it H}$_{o}$= $\frac{\phi _{o}}{\pi ^{2}\lambda
\xi }$, and ${\it \phi }_{o}$ is the flux quantum. For ${\it \lambda }_{ab}$
= 1400\AA\ and ${\it \xi }_{ab}$ = 20 \AA , which are typical values, {\it H}%
$_{o}\approx $ 8000 Oe. For measurements in a field of amplitude 300 Oe,
this yields $\frac{H}{H_{o}}$ = 0.0375, and requires that T $\lesssim $ 10K.

To measure {\it m}$_{T}$, an AC magnetic field is applied in the {\it a-b}
plane of a mm size single crystal of YBCO. The field is modulated at 12Hz,
and the transverse magnetic moment at 36Hz is detected by the transverse
coil of a Quantum Design SQUID susceptometer. The signal is expected at the
third harmonic because {\it m}$_{T}$ $\sim $ {\it sgn(H)H}$^{2}$. The sample
is maintained at 2.5K with continuous cooling throughout the measurements.
The superconducting magnet is operated with its persistent-current switch
open, and the AC magnetic field is generated by driving a very low
distortion (%
\mbox{$<$}%
-110dB in higher harmonics) and low noise AC current through the magnet
coil, with a maximum amplitude of 300 Oe. The sample is placed in the center
of the magnet coil and aligned optimally with the transverse flux detection
coils. The analog output from the transverse SQUID is fed directly to a
phase sensitive detection system. The large 12Hz background from the
fringing field of the magnet and geometric demagnetization fields from the
sample are rejected by a high-pass filter section. The signal at 36Hz is
then detected with a phase sensitive detector locked to {\em 3f} of the
current generator. The sample is rotated {\em in situ} between measurements
in steps of 6$^{o}$ using a custom built sample holder that allows for {\em %
in situ }detection of the angular position of the sample at low
temperatures. This system has been described in detail elsewhere \cite{rsi}.
It allows for precision in angular positioning of about 0.1$^{o}$ between
steps and accuracy of better than 1$^{o}$ in an entire rotation. The sample
holder and associated angular detection systems are designed to provide
minimal magnetic background.

The detection setup has been calibrated and tested by `simulating' a real
magnetic moment in an environment identical to that of the actual
experiment. This is done by running an AC current through a copper coil
mounted on the sample stage at a frequency {\em 3f}, referred to the
frequency {\it f }=12 Hz of the oscillating magnetic field, which has an
amplitude of 300 Oe. The AC moment of 1.5 x 10$^{-8}$ emu rms amplitude at 
{\em 3f} is detected in the presence of the large background signal at
frequency {\em f} due to the fringing field of the magnet. This calibration
was done at 2.5K, with just the bare sample holder. The measured amplitude
is shown in Fig.1 as a function of the angular orientation of the coil. The
angular Fourier transform of this data (inset) shows the amplitude at
angular period {\it 2}${\it \pi }$. The `noise floor' in the Fourier
transform of the calibration signal at higher angular frequencies
corresponds to an amplitude less than 5 x 10$^{-10}$ emu.

The angular dependence of the NLME has been measured in a number of
untwinned single crystals of YBCO, including a very high quality crystal
(UWHC), grown in a YSZ crucible by the Argonne group. This crystal, which is
99.95\% pure, has been cut into a disk shape with diameter 1.5mm, and is 67$%
\mu m${\em \ }thick. The high quality of the crystal is borne out by a
rocking curve with full width at half maximum of  0.08$^{0}$ for the 006
peak of a high resolution x-ray scan, and there were no twins observable in
repeated area scans over different parts of the crystal. The superconducting
transition of this crystal has an onset of 93K and a width of less than 2.5K
as measured magnetically with a 10 Oe field applied in the {\it a-b} plane.
The field of first flux entry was measured to be about 300 Oe in the {\it a-b%
} plane at 2.5K.

Measurements of {\em m}$_{T}$ are shown in Fig.2 as a function of the
angular orientation of the sample with respect to the applied magnetic
field, with the{\em \ `a'} direction being initially oriented along the
field. The sine Fourier transform of the data is shown in the inset. For a
pure {\em d-wave} order parameter, we are interested in the angular period 
{\it 2}${\it \pi }${\it /4} component of {\em m}$_{T},$ the predicted value
for which is 1.7 x 10$^{-9}$emu according to the calculations in Ref.\cite
{zv1}. This component is clearly in the noise of the data, below 5 x 10$%
^{-10}$ emu. This 'noise' is in part due to residual trapped flux, and also
due to the noise floor of the measurement apparatus, as has been verified in
repeated measurements. This imposes an {\it upper bound} on the size of the
nonlinear Meissner effect, which if there, is less than 30\% of the
predicted value. Identical measurements have also been carried out on a very
high quality rectangular crystal (UBCyca) grown by the UBC group in a BaSZ
crucible\cite{hardy98}. It has the same surface area as our disk shaped
crystal (UWHC) and the results were essentially the same.

This null result has to be examined in the light of measurements of the
penetration depth. Scrutiny of data on the temperature dependence of the
penetration depth ${\it \lambda }${\it (T)} indicates that there is always
some curvature away from the pure {\it d-wave} result at the lowest
temperatures. This is found in all published data on the cleanest bulk
single crystals \cite{hardy93,hardy98}, and has been attributed to the
presence of unitary scattering centers, that do not affect T$_{c}\cite
{zn,unitary}$. However, even with the concentration of unitary scatterers
needed to produce the required curvature, the nonlinear transverse magnetic
moment effect may only be reduced to about 90\% of the full value \cite{ys},
and our data indicates a much stronger suppression. This same data for ${\it %
\lambda }${\it (T) }for UBCyca \cite{chris} can be fit at the lowest
temperatures by a model which contains quasinodes, i.e., where the $\frac{%
\Delta _{\min }}{\Delta _{o}}$ is 2.5\%, ${\it \Delta }_{\min }$ being the
residual gap in the nodal direction. Such a fit is significantly better at
the lowest temperatures than the fit for a pure {\it d-wave} order
parameter. The suppression of the nonlinear transverse magnetic moment in
all our measurements is consistent with $\frac{\Delta _{\min }}{\Delta _{o}}$
between 2 - 3\%. In the event that the ${\it \lambda }${\it (T) }and
photoemision results can be explained by something other than a minimum in
the gap, our experiment is also consistent with a{\em \ d+s} order parameter
where $\Delta _{s}$ 
\mbox{$>$}%
$\Delta _{d}.$

Recently, other experiments that have attempted \cite{giannetta,hardy} to
see the NLME in measurements of the field dependence of ${\it \lambda }$ at
low temperatures, but have not obtained conclusive results. According to
theory\cite{ys}, for a pure {\em d-wave} order parameter, ${\em \lambda }$
should vary linearly with {\em H} at the lowest temperatures where $\frac{T}{%
{\it \Delta }_{0}}$ 
\mbox{$<$}%
$\frac{H}{H_{o}}.$ These most recent measurements were done at a very high
level of sensitivity, about 100 times better than an earlier effort \cite
{maeda}. In one of these experiments done at UIUC \cite{giannetta}, a
significant field dependence in ${\em \lambda }$ was observed, but this was
attributed to trapped flux as the field dependence was closer to $\sqrt{H}$%
and the temperature dependence was not as predicted by theory. The
experimental effort at UBC\cite{hardy} observed a field dependence in ${\em %
\lambda },$ and also measured the effect as a function of the in-plane
angular orientation of the applied field. However, they report that the
field dependence measured at $\pm $ 45$^{o}$ to the crystal axes were not
identical as would have been expected from theory, and the temperature and
field dependence of the effect could not be understood within the current
picture for the NLME \cite{ys}.

Earlier attempts at measuring the angular dependence of the nonlinear
transverse magnetic moment using a DC technique \cite{buan} were less
sensitive by more than two orders of magnitude than the present work due to
the very large linear background in a DC measurement, and the analysis
overestimated the size of the effect significantly. In our experiments,
trapped flux may produce a signal, but the angular dependence of this effect
has the largest amplitude at angular period {\it 2}${\it \pi }$, and is
easily distinguished from the period {\it 2}${\it \pi }${\it /4 } modulation
for which we are searching. The angular modulation arising from the
geometric demagnetization factor is linear in the field below the field of
first flux entry, and is minimized by looking at only the nonlinear
component which is extracted directly in our work. Thus, artifacts due to
trapped flux and geometry are both minimized by our technique.

There have been some calculations of the effects of nonlocal electrodynamics 
\cite{leggett,hirschfeld}. These are motivated by the idea that in a BCS
like picture, when the gap ${\it \Delta }_{k}$ goes to zero, then ${\it \xi }
$ = $\hbar ${\it v}$_{F}$/${\it \pi \Delta }_{k}$ diverges, and one might no
longer be in the local limit where ${\it \xi }$ 
\mbox{$<$}%
{\it \ }${\it \lambda }$ . However, it turns out that these considerations
are relevant only when $\widehat{{\it H}}${\it \ 
\mbox{$\vert$}%
\mbox{$\vert$}%
}$\widehat{{\it c}}$. In our experiment, $\widehat{{\it H}}$ is applied in
the {\it a-b} plane, and the volume within a depth ${\it \lambda }$ of the 
{\it a-b} plane crystal face that is responsible for the nonlinear Meissner
effect is not affected by nonlocal effects. Even if we consider a `weakly
3D' case \cite{hirschfeld}, the effect is at a field scale of about 20 Oe,
far too small to suppress the NLME, whose characteristic field in our
measurements is 300 Oe.

In summary, in view of the results of this paper, and in light of the
evidence for a predominantly {\it d-wave} order parameter \cite{squids} from
other experiments, we suggest that the order parameter in YBCO may still be 
{\it d-wave} like, but have quasinodes instead of line nodes, with $\frac{%
\Delta _{\min }}{\Delta _{o}}$ between 2 - 3\%. Measurements of the
penetration depth to lower temperatures or photoemisssion experiments at
higher resolution may confirm this.

We'd like to acknowledge R.Giannetta, J.Buan, B.P.Stojkovic, R.Klemm, and
D.E.Grupp for helpful discussions. We are also extremely grateful to
C.Bidinosti, D.Bonn, R.Liang, W.Hardy of the UBC group for letting us
confirm our results with one of their very high quality crystals of YBCO
(UBCyca), and also for freely sharing their data on $\lambda (T)${\em \ }and 
{\em \ }$\lambda (H)$ for this and other crystals with us before
publication. The early stages of this work were supported in part by the
AFOSR under grant F49620-96-0043. One of us (A.B.) would like to acknowledge
Foster Wheeler and Graduate Dissertation Fellowships from the Graduate
School of the University of Minnesota.


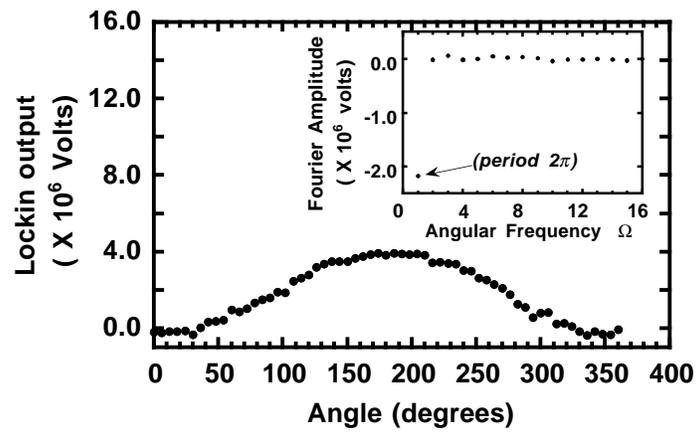
\begin{figure}[tbp]
\caption{Sensitivity of the measurement apparatus to a magnetic moment of
1.5 x 10$^{-8}$ emu at {\em 3f}. The amplitude at period {\em \ }${\em 2\pi }
$ is shown in the cosine Fourier transform (inset) and is 2.2$\mu $V. }
\label{Fig.1}
\end{figure}

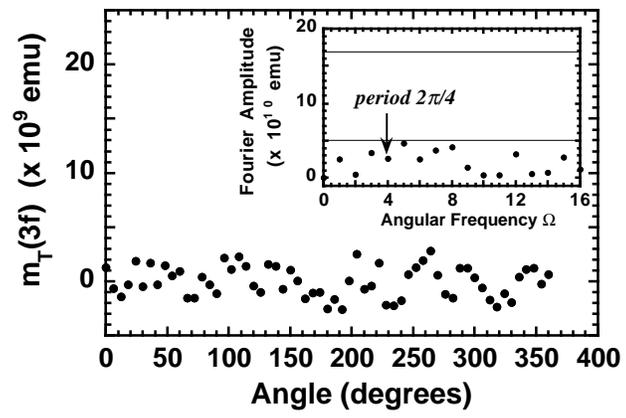
\begin{figure}[tbp]
\caption{{\em m}$_{T}${\em (3f)} as a function of angle with with respect to
applied magnetic field for UWHC YBCO sample. The initial orientation is with
the `{\em a}' direction parallel to the applied magnetic field. The {\em %
absolute} value of the sine Fourier amplitudes are plotted in the inset. The
expected amplitude at angular period {\em 2}${\em \pi }${\em /4} modulation
for pure {\em d-wave} is shown as a horizontal line at 17 x 10$^{-10}$ emu.
The level predicted for ${\em \Delta }_{\min }$/${\em \Delta }_{o}$= 2.5\%
is shown by the horizontal line at 5 x 10$^{-10}$ emu, near the noise floor.}
\label{Fig.2}
\end{figure}


\end{document}